# Mitigating Interpretation Bias in Rock Records with Large Language Models: Insights from Paleoenvironmental Analysis

**Luoqi Wang[1,2], Haipeng Li[3,4]\*, Linshu Hu[1,2]\*, Jiarui Cai[1,2] and Zhenhong Du[1,2]**

[1]*School of Earth Sciences, Zhejiang University, Hangzhou 310027, China;*

[2]*Zhejiang Provincial Key Laboratory of Geographic Information Science, Hangzhou 310028, China;*

[3] *Deep-time Digital Earth Research Center of Excellence (Suzhou), Kunshan, 215347, China;*

[4] *Deep-time Digital Earth Research Center of Excellence (Hangzhou), Hangzhou, 311121, China;*

*\* Co-Correspondence to:* Linshu Hu (hulinshu1010@zju.edu.cn) & Haipeng Li (li.haipeng@ddeworld.org)

**ABSTRACT**

The reconstruction of Earth's history faces significant challenges due to the nonunique interpretations often derived from rock records. The problem has long been recognized but there are no systematic solutions in practice. This study introduces an innovative approach that leverages Large Language Models (LLMs) along with retrieval augmented generation and real-time search capabilities to counteract interpretation biases, thereby enhancing the accuracy and reliability of geological analyses. By applying this framework to sedimentology and paleogeography, we demonstrate its effectiveness in mitigating interpretations biases through the generation and evaluation of multiple hypotheses for the same data, which can effectively reduce human bias. Our research illuminates the transformative potential of LLMs in refining paleoenvironmental studies and extends their applicability across various sub-disciplines of Earth sciences, enabling a deeper and more accurate depiction of Earth's evolution.

**INTRODUCTION**

The Earth's rock records hold a wealth of information about the planet's history, climate, and the life forms (Boucot et al., 2013; Talent, 2012; Torsvik & Cocks, 2016) that have inhabited it. Geologists interpret these records to reconstruct past environments and understand the processes that have shaped our planet (Bridge & Demicco, 2008; Keighley, 2013). However, interpretation is often fraught with challenges due to the underdetermination of theory by data and human bias.

In Earth science, underdetermination manifests when data do not provide enough information to conclusively support one interpretation over another (Kleinhans et al., 2010; Turner, 2005). The phenomenon of underdetermination has long been recognized (Gilbert, 1886), with various other terms like nonuniqueness (Kim & Ivanov, 2014; Oreskes et al., 1994), equifinality (Chorley, 1962; Nicholas & Quine, 2010), and ambiguity (Hanneson, 2022; Roy, 1962) being used to convey the same idea. Well-known examples of nonunique interpretations include controls on stratigraphic patterns (Burgess & Prince, 2015; Madof et al., 2016), fault-plane solutions (Brinkman et al., 2021), causes for seismic velocity anomaly and tomography (Rawlinson et al., 2014), and implications of δ18O of benthic foraminifera (Boer et al., 2010), to name just a few.

Additionally, we are all biased in terms of education and prior experience, and this affect how we interpret the rock records (Li & Plink-Björklund, 2019). For instance, the early debate on Neptunism and Plutonism is in part due to different field experience of Abraham G. Werner and James Hutton, where Werner's fieldwork is mostly confined to the Erzgebirge and neighboring areas, while Hutton was exposed to the Midland Valley of Scotland that contains numerous basic sills from which he made the critical observations that convinced him of the plutonic origin of rocks (Hallam, 1990). Other examples relate to the various forms of cognitive biases (Tversky & Kahneman, 1974). For example, the availability of different theoretical frameworks across disciplines is a key driver of disciplinary bias, leading to divergent interpretations of the same underlying information. A seismologist is apt to interpret seismic velocity anomalies in terms of temperature variations, while a petrologist is more likely to interpret the anomalies caused by variations in mineralogy, crystal orientation or partial melt content (Anderson, 2007).

To effectively counteract the underdetermination, T. C. Chamberlin (1890) proposed the method of multiple working hypotheses, which is perhaps the best solution in theory. Some of the prominent success of this method include J. T. Wilson's discovery of transform faults, which laid the foundation of plate tectonics (Laudan, 1980). However, due to our own limitation in training, experiences, and psychological reasons, the method is more widely advocated (Chamberlin, 1890; Dalrymple & James, 2010; Platt, 1964) than practiced (Yanco et al., 2020), even for Chamberlin himself in the case of Chamberlin-Moulton planetesimal theory (Brush, 1978). The exponential increase of publications (Bornmann et al., 2021) further worsen the issue; it is becoming exceedingly difficult to keep up with publications, while at the same time getting a good grasp on the current and existing literature serves as an importance source for ideas and hypotheses, especially considering the fact that Earth science relies heavily on case-based reasoning, the situation seems a little bit hopeless if we cannot keep up with the ever-growing cases.

Methods for dealing with human bias in decision have also been proposed, namely through the application of statistical models and algorithms (Green & Chen, 2019; Meehl, 1954). Statistical models and algorithms have been found to consistently outperform experts in many tasks (Dawes et al., 1989; Grove et al., 2000), and biases in algorithms can be corrected much easily compared to changing the human decision-making process. This is because algorithms are often more transparent and explainable than human decision-making processes, and they can be systematically updated and corrected as new data becomes available or as biases are identified. There have been attempts to create algorithms and systems to facilitate paleoenvironmental reconstruction at the outcrop and basin scale based on knowledge engineering (Miller, 1993; H. Wang et al., 2023; Zhang et al., 2023), and they have achieved certain success. However, there are three major challenges. One is the construction of a knowledge base, which often takes a long time and lots of effort; in addition, regularly updating the knowledge base is also no small task. The second is that such a knowledge-based system often requires semantic parsing, which can be difficult for lots of semi-structured or unstructured data in papers and databases such as GeoLexicon (Du et al., 2023).

The rapid development of large language models (LLMs) (Kaddour et al., 2023; Kojima et al., 2022) has brought new opportunities for solving the underdetermination and bias problems in paleoenvironmental reconstruction. Such models, such as ChatGPT and Claude, are trained on a large corpus of text covering essential all areas of human knowledge (Jiang et al., 2023). At the same time, they can be quite effective for understanding natural language queries (Zhao et al., 2023), reasoning about tasks (Wei et al., 2023), and generating ideas and hypotheses (Wei et al., 2022). A preliminary test suggests large language models can be effective in paleoenvironmental reconstruction. When given proper prompt and a description of the field observation, ChatGPT can give the depositional processes, list corresponding possible environments ranked by their probability, modern analogs, and references.

However, there are two main problems need to be addressed before we can use it for paleoenvironmental interpretation. The first problem is what has been called AI hallucination (Huang et al., 2023); put it another way, how can we minimize the proportion of content that is made up by LLMs? The other problem is how to make the LLMs to keep up with the latest research findings when they are initially trained on text that typically has a cut-off date, such as 2022.

To get the most out of LLMs in the application of paleoenvironmental reconstruction, our solution is to use chain-of-thought (Wei et al., 2023), proper prompt based on our aim (Beurer-Kellner et al., 2023), retrieval-augmented generation (Gao et al., 2024), and online search on top of GPT4 API.

## DATA AND METHOD

### Data and Pre-processing

The quest to reconstruct Earth's history is a complex and intricate endeavor that has long captivated geologists and Earth scientists (Milanovsky, 2007). The rock record, our primary source of information about the Earth's past, is a treasure trove of data that, when interpreted correctly, can reveal the planet's dynamic history (Duarte et al., 2021; Hedberg, 1965). However, this scientific methodology is not without its challenges. Nonunique interpretations often arise from the rock record, complicating our understanding of the geological history of the Earth (Hamilton, 2019; Saltus & Blakely, 2011). Meanwhile, interpreting based on incomplete evidence is actually a rule in geology, rather than an exception (Frodeman, 1995). More reliable research materials can support more robust and truthful theoretical hypotheses, just as the theory of Sea-Floor Spreading is formed and supported by the integration of a large number of observations and hypotheses, and becomes a further development of the Continental Drift (Heirtzler, 1968). Based on this background, we hope to reduce bias in information acquisition and evidence interpretation by integrating relevant data materials and knowledge in the geological field to assist in hypothesis generation and evaluation.

For the above purpose, we searched and obtained thousands of literatures in the field of paleogeography and sedimentology to obtain their meta information, as the essence of domain knowledge. Our metadata dataset contains basic information about articles, including DOI, abstract, title, authors, publication date, etc., for reference. For the obtained metadata, the main focus is on processing the abstract fields. We segmented the abstract text into paragraphs and sentences, and obtained query tables for paragraphs and statements. We introduced a sentence transformer model (sentence-transformers/all-MiniLM-L6-v2) to maps these paragraphs and statements to a 384-dimensional dense vector space, and stored them in a vector database for retrieving.

**Expert Question and Answer System**

The core abilities of large language models (LLMs), such as ChatGPT, encompass language generation and emergence ability, enabling it to exhibit emergent behavior while generating coherent and contextually relevant text (Ouyang et al., 2022; Wei et al., 2022). Since the advent of ChatGPT, LLMs show great potential in assisting scientific research. The application of LLMs in disciplinary vertical fields, such as law, medical, chemistry and economics, is constantly evolving (Cui et al., 2023; Dan et al., 2023; L. Wang et al., 2023; N. Wang et al., 2023; Xiong et al., 2023). The ability of LLMs to integrate, understand, and apply scientific literatures in reasoning and interpretation has been widely reflected in existing research (Jablonka et al., 2023; Miret & Krishnan, 2024; Zheng et al., 2023). Their ability to summarize and generalize data that far exceeds human capabilities allows them to explore and present diverse perspectives, solutions, and creative ideas. Models such as K2 and GeoGalactica demonstrate the potential for conducting knowledge reasoning and thought generation tasks in the field of geology to discover potential undiscovered relationships (Deng et al., 2023; Lin et al., 2023). This, in turn, brings new insights and possibilities for problem-solving and scientific research.

However, despite its impressive capabilities, LLMs may encounter challenges such as hallucination, which means the information they generate is nonsensical or unfaithful (Ji et al., 2023). This issue arises due to the models' ability to generate text only based on patterns and associations learned from vast amounts of training data, without necessarily verifying the authenticity of the associations (Huang et al., 2023; Ji et al., 2023). Furthermore, while LLMs excel in generating general text, their performance in specialized or professional domains may be insufficient. This limitation stems from their lack of domain-specific knowledge and expertise, making them less reliable for tasks that require in-depth understanding and specialized knowledge (Kaddour et al., 2023b; Zhao et al., 2023).

To enhance the ability of LLMs in professional fields and meet the demands of solving the problem of scientific research bias caused by insufficient knowledge acquisition, simultaneously addressing hallucination issues, several solutions can be considered. These include fine-tuning the models on domain-specific data, incorporating external knowledge sources to improve accuracy, and developing specialized models tailored to specific professional domains (Ciucă et al., 2023; Huang et al., 2023; Ouyang et al., 2022). Additionally, implementing mechanisms for fact-checking and verification can help mitigate the issue of hallucination and ensure the generation of reliable and trustworthy information (Pan et al., 2023). In the subsequent discussion, we outline our methodology, which involves configuring specialized knowledge bases along with real-time external knowledge sources, coupled with the rethink mechanism, aimed at alleviating hallucination and bolstering the system's proficiency within professional domains.

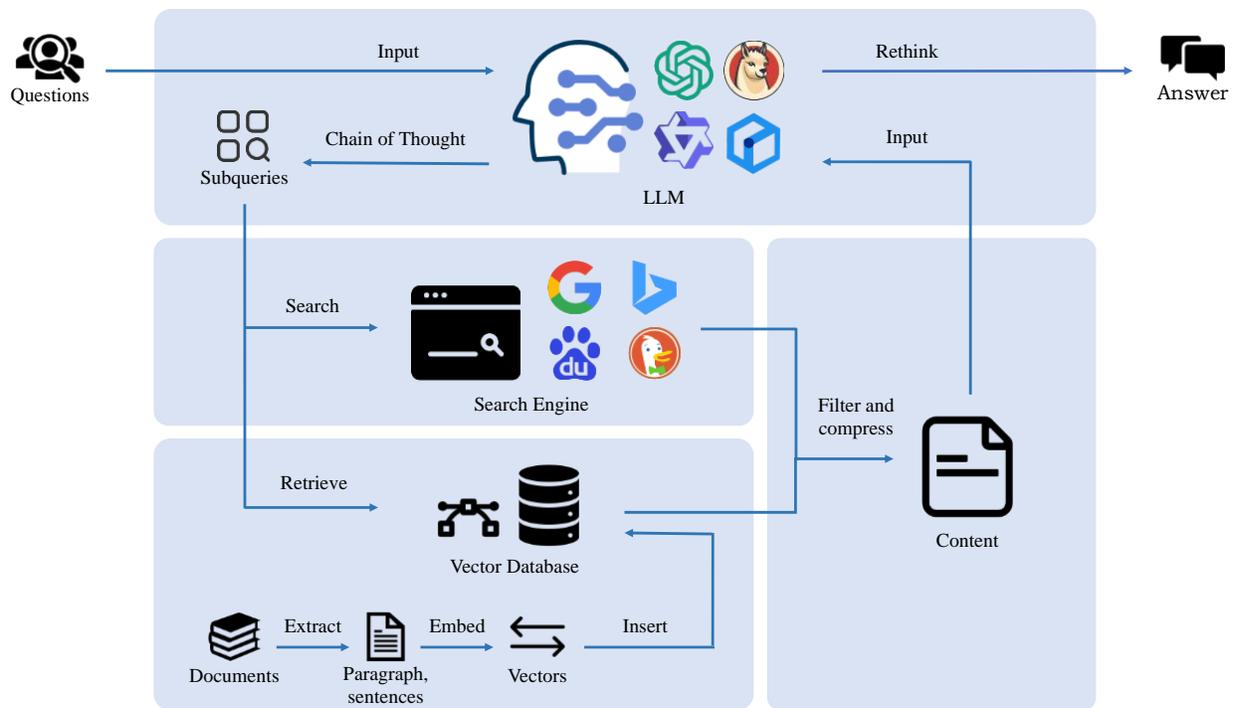

Figure 1. The foundational architecture of the expert question and answer system

As illustrated in Figure 1, we have developed an expert question and answer (QA) system leveraging LLMs. This system is adept at responding to user queries by accessing and assimilating information from domain-specific knowledge repositories as well as dynamic, real-time external data sources. This augmentation significantly bolsters the analytical and cognitive capabilities of LLMs within targeted domains, thereby facilitating the provision of expert-level responses. The functional architecture of the expert system can be delineated into the following three modules: (1) LLM module, (2) data retrieval and retrieval module, and (3) generation module.

**LLM module**

The LLM module functions as the central processing unit of the expert system, receiving diverse inquiries from users and generating responses accordingly. This module enhances the system's cognitive capabilities by leveraging the services of language models such as GPT-4.

The core capability of the expert question and answer system is built on the generation and emergence capabilities of LLMs. The generation ability of LLMs is an important feature that enables the expert system to generate text with coherence and contextual relevance. This ability arises from their training on extensive corpora of text data, allowing LLMs to learn patterns, linguistic structures, and semantic relationships (Huang et al., 2023; Kojima et al., 2023; Naveed et al., 2023; Zhao et al., 2023). The concept of emergence refers to the phenomenon where simple components interact and reach a certain scale, resulting in complex behaviors and patterns. In the case of LLMs, emergence is observed as they generate text that exhibits a deep understanding of language, despite not being explicitly programmed for specific tasks. The emergence ability of LLMs is based on the number of model parameters and the amount of computation, and sourced from the vast and diverse text data they are trained on, which provides them with a rich knowledge base to draw upon (Arora & Goyal, 2023; Wei et al., 2022). To enhance the efficacy of LLMs in generating outputs aligned with our expectations and specific problem contexts, the utilization of the "chain of thought" (CoT) technique is prevalent within the system. This approach significantly enhances the multi-step reasoning capabilities of LLMs. Through the strategic insertion of prompts like "step by step," LLMs autonomously establish logical connections among diverse concepts, fostering the construction of coherent and contextually relevant responses (Wei et al., 2022; Yao et al., 2023).

In our expert system, we utilize the CoT to replicate expert cognitive processes for question analysis. Through this approach, user-inputted questions are systematically deconstructed into manageable sub-queries, facilitating a structured examination. The application of CoT is realized through prompt engineering, a method wherein researchers and developers strategically craft input prompts. These prompts are tailored considering linguistic nuances, contextual factors, and structural elements, while also providing explicit instructions or cues. This meticulous design aims to guide the model towards generating responses that adhere to specific criteria or objectives (Beurer-Kellner et al., 2023; Jiang et al., 2023; Zhao et al., 2023). Prompt engineering encompasses diverse strategies, such as incorporating explicit instructions, specifying desired output formats, or integrating supplementary contextual information. By adopting this approach, LLMs are trained to emulate expert reasoning, engaging in a systematic, step-by-step thought process. Through ingeniously crafted prompts, LLMs can develop coherent thought chains, gradually unraveling complex problems.

**Data search and retrieval module**

The data search and retrieval module provide knowledge support for LLM module. Upon decomposition of user queries into subqueries by LLMs, these subqueries are sequentially inputted into this module. Through the retrieval of information pertinent to each subquery, the module facilitates the provision of highly correlated content, thereby aiding LLMs in problem-solving and reevaluation.

The retrieval module described herein comprises two primary data sources: a vector database housing professional data and search engines for accessing public fundamental data. The vector database stores disciplinary literature, serving as a repository for professional data. Through the extraction and embedding of literature data, textual content is transformed into vectors and stored in the database, facilitating efficient retrieval and computation (Pan et al., 2023). Subsequent retrieval from the vector database involves employing the vector similarity algorithm, wherein the Euclidean Distance metric is utilized for calculating vector similarity. Alternatively, employing Inner Product (IP) or other vector similarity algorithms may yield comparable or superior results. Upon querying the vector database, the algorithm searches for records containing the nearest N sentence or paragraph vectors within the vector knowledge base. Concurrently, search engines are engaged to process subqueries by leveraging common Internet search services, with Google serving as the primary example. Other search engines such as Bing and Baidu can serve as viable alternatives. Information retrieved from both databases and search engines undergoes embedding via a sentence transformer model. Subsequently, autocorrelation calculations are applied to filter out items exhibiting the highest autocorrelation. A notable challenge with information exhibiting high autocorrelation is the typically extended length of their content. Addressing this challenge, contextual compression is employed to retrieve and compress the context of a given content, eliminating extraneous content while retaining the most pertinent information relevant to the query problem. Finally, the returned information is standardized to include three essential fields: source, title, and content.

From the perspective of the respective functions of search engines and vector databases in the data search and retrieval module, search engines play a vital role in furnishing essential reasoning materials for LLMs, thus establishing a foundational basis for expert system capabilities. Simultaneously, scientific research-based knowledge stored in vector databases enriches and supplements the expertise of expert systems, elevating their proficiency within specialized domains.

**Generation module**

The generation module serves as the primary functional component within this system, receiving outputs from the data retrieval module and invoking the LLM module to summarize and condense information, ultimately generating conclusions.

Retrieval-augmented generation (RAG) represents an innovative approach amalgamating retrieval-based methods with generative models, typically Language Models (LMs), to enrich the quality and relevance of generated text (Ciucă et al., 2023; Gao et al., 2024). This article's expert question and answer system embodies the concept of RAG. In this paradigm, LLMs retrieve input questions (segmented into subqueries via CoT) from the data retrieval module to acquire information for enhancing generation. Subsequently, LLMs reconsider both the acquired information and original questions. During the reconsideration phase, prompt engineering, leveraging task description, input data, contextual information, and prompt style, enriches and standardizes LLM output (Zhao et al., 2023), enabling LLM to produce answers meeting predefined standards, akin to those of domain experts or proficient students.

In the realm of RAG research, advanced and modular RAG methodologies are evolving from the original or naive RAG approach (Gao et al., 2024). Advanced RAG incorporates additional processing stages pre- and post-retrieval. Pre-retrieval processing concentrates on optimizing data indexing via various methods, including data granularity refinement, index structure optimization, and metadata incorporation to enhance retrieval content quality. Post-retrieval processing involves reranking and prompt compression, with embedding playing a crucial role. Optimization strategies encompass fine-tuning embedding or employing dynamic embedding methods. Embracing the design principles of Advanced RAG, we enhance the pre-retrieval stage with query planning and expansion, while bolstering post-retrieval stages with reranking and summarization techniques.

**RESULT AND DISCUSSION**

**System Performance**

We developed an expert question and answer (Q&A) system accessible through a user-friendly web application. Users input their queries on the web page and specify the relevant domain, prompting the system to generate structured responses. The interface offers real-time insights into the system's cognitive process and presents the finalized answers, which users can export in Portable Document Format (PDF).

Our system draws upon a comprehensive knowledge base curated from metadata extracted from a vast corpus of sedimentology literature. In selecting the LLM, we opted for GPT-4 due to its superior performance characteristics.

To assess the system's efficacy, we conducted a human evaluation focusing on its utility by engaging domain experts. This evaluation involved posing various questions to the system and qualitatively assessing the appropriateness of its responses. An illustrative example is provided in the Appendix. Feedback from experts in paleogeography and sedimentology indicates that the system's responses closely approximate those of proficient students or researchers in the respective fields.

The expert system showcases its capacity to generalize and comprehend extensive geological knowledge, facilitating multi-perspective analysis of geological phenomena and hypotheses. For instance, when tasked with proposing hypotheses for specific geological phenomena, the system generates multiple hypotheses accompanied by relevant analyses, ranking them based on plausibility. These results generally align with the consensus among domain experts.

## Figure 2. System operation example

**Left panel — User Question and Process:**

**Question:** What are the possible causes for the earliest Miocene shift from a planetary-wind-dominant type to a monsoon-dominant type in China? List **different hypotheses ranked** by their probability and **explain** why.

**Paleogeography Expert Q&A System**

**Process**

🤔 Thinking about research questions for the task...
🔍 Running research for 'What are the possible causes for the earliest Miocene shift from a planetary-wind-dominant type to a monsoon-dominant type in China? List different hypotheses ranked by their probability and explain why.'...
🌐 Paleogeography Agent
💬 I will conduct my research based on the **following queries**: ['causes of Miocene climatic shift to monsoon dominance in China', 'hypotheses on Miocene planetary wind to monsoon transition probability', 'explanations for early Miocene monsoon intensification in China 2024', 'What are the possible causes for the earliest Miocene shift from a planetary-wind-dominant type to a monsoon-dominant type in China? List different hypotheses ranked by their probability and explain why.']...
🔍 Running research for 'causes of Miocene climatic shift to monsoon dominance in China' in **internet**...
✅ Adding source url to research: https://www.sciencedirect.com/science/article/abs/pii/S092181812300320X
✅ Adding source url to research: https://en.wikipedia.org/wiki/Miocene
...
🤔 Researching for relevant information...
🌐 Summarizing: causes of Miocene climatic shift to monsoon dominance in China
📄 Source: https://en.wikipedia.org/wiki/Miocene Title: Miocene Content: The global trend was towards increasing aridity caused primarily by global cooling reducing the ability of the atmosphere to absorb moisture,[19] particularly after 7 to 8 million years ago.[20] Uplift of East Africa in the late Miocene was partly responsible for the shrinking of tropical rain forests in that region,[21] and Australia got drier as it entered a zone of low rainfall in the Late Miocene…
🔍 Running research for 'causes of Miocene climatic shift to monsoon dominance in China' in **DB**…
🌐 Summarizing: causes of Miocene climatic shift to monsoon dominance in China
📄 Source: https://doi.org/10.1016/j.sedgeo.2021.106030 Title: Damp-to-dry aeolian systems: Sedimentology, climate forcing, and aeolian accumulation in the Late Cretaceous Liyou Basin, South China Content: Aeolian desert systems were widely distributed in South China during the Late Cretaceous. These systems developed under the control of subtropical highs and a monsoon climate. They reveal the evolution of palaeoclimate and palaeogeography and the regional palaeowind pattern in the South China interior during this period. Based on the analyses of sedimentology and facies architecture, this study examines and reassesses the aeolian succession of the Upper Cretaceous Daijiaping Formation in the southern Liyou Basin of southeast China. ...
✏️ Writing: What are the possible causes for the earliest Miocene shift from a planetary-wind-dominant type to a monsoon-dominant type in China? List different hypotheses ranked by their probability and explain why....

**Right panel — Answer:**

**Paleogeography Expert Q&A System**

**Answer**

**The Shift to a Monsoon-Dominant Climate in Early Miocene China: Evaluating Hypotheses**

**Introduction**
The transition from a planetary-wind-dominant climate to a monsoon-dominant climate in China during the Early Miocene epoch represents a significant climatic shift with profound implications for the region's paleoenvironment and sedimentology. This report examines the potential causes of this climatic transformation, evaluating various hypotheses based on their likelihood and supporting evidence. The Miocene epoch, spanning from approximately 23 to 5 million years ago, was a period of considerable climatic change, marked by fluctuations in global temperatures and atmospheric compositions that influenced precipitation patterns and vegetation distribution **(Miao et al., 2019)**.

**Hypotheses for the Climatic Shift**
**Uplift of the Tibetan Plateau**
Probability: **High**
The most widely supported hypothesis for the shift to a monsoon-dominant climate in Early Miocene China is the uplift of the Tibetan Plateau. The collision of the Indian subcontinent with Asia around 50 million years ago initiated the plateau's rise, which intensified during the Miocene…

**Global Cooling and Antarctic Ice Sheet Expansion**
Probability: **Moderate**
...

**Tectonic Activity and Regional Uplift**
Probability: **Moderate to High**
...

**Changes in Ocean Circulation and Sea Surface Temperatures**
Probability: **Low to Moderate**
...

**Discussion**
The uplift of the Tibetan Plateau stands as the most probable cause for the shift to a monsoon-dominant climate in Early Miocene China. This hypothesis is supported by the temporal correlation between the plateau's uplift and the intensification of the monsoon, as well as the physical mechanism by which the plateau could influence atmospheric circulation…

**Conclusion**
In conclusion, the uplift of the Tibetan Plateau is the most probable cause for the shift to a monsoon-dominant climate in Early Miocene China, with tectonic activity and regional uplift also playing significant roles. Global cooling and changes in ocean circulation provide additional context for understanding the complex interplay of factors that contributed to this climatic transformation. Further research, particularly involving deep ocean sampling and high-resolution paleoclimatic records, is necessary to refine our understanding of these processes and their relative contributions to the development of the monsoon system.

**References**
Middle Miocene disruption. (n.d.). Retrieved from https://en.wikipedia.org/wiki/Miocene
Miao, Y., Wu, F., Warny, S., Fang, X., Lu, H., Fu, B., … Shi, P. (2019). Miocene fire intensification linked to continuous aridification on the Tibetan Plateau. Geology, 47(4), 303–307. https://doi.org/10.1130/G45720.1
...

Figure 2. System operation example

Figure 2 shows an operation example of the paleogeography expert question and answer system. Enter the question into the expert question answering system: "What are the possible causes for the earliest Miocene shift from a planetary-wind-dominant type to a monsoon-dominant type in China? List different hypotheses ranked by their probability and explain why."

As shown in Figure 2 (some repeated processes are omitted), the paleogeography expert question and answer system outputs the process of its thinking, making it easier for users to understand how it operates. In the case, the expert system first defines itself as an expert agent in the field of paleogeography, and begins to think about user input question. In system output, it can be seen that the user input question has been split into four queries: 'causes of Miocene climatic shift to monsoon dominance in China', 'hypotheses on Miocene planetary wind to monsoon transition probability', 'explanations for early Miocene monsoon intensification in China 2024', 'What are the possible causes for the earliest Miocene shift from a planetary-wind-dominant type to a monsoon-dominant type in China? List different hypotheses ranked by their probability and explain why.'. Then, the expert system starts to query the four questions in turn on the Internet, obtain relevant links, and organize the content of the link page into the format of source, title, and content triplets. Similarly, the expert system then conducted sequential queries and content organization in the database. The final retrieved content is used for generating the final research report.

The final generated report is shown on the right side of Figure 2 (with some paragraphs omitted). The format of this research report is very organized and basically meets the requirements of scientific research texts, including a title, outline structure, and paragraph content. The paragraphs include introduction, possibilities and explanations of various hypotheses, discussions, conclusions, and references. The statement in the report has citation sources, such as a viewpoint in the introduction, "The Miocene epoch, spanning from approximately 23 to 5 million years ago, was a period of considerable climatic change, marked by fluctuations in global temperatures and atmospheric compositions that influenced precipitation patterns and vegetation distribution", that cited Miao's article, Miocene fire intensification linked to continuous aridification on the Tibetan Plateau. This indicates that our system has indeed enhanced its generated reliability through RAG.

**Ablation Experiment**

To demonstrate the effectiveness of the retrieval enhanced design in the expert system in generating of LLMs, we conducted an ablation experiment. We separately removed the support of vector database retrieval and search engine search to compare with the complete expert question and answer system. We have invited experts in the field to provide more than ten professional questions for experimentation, all of which require multiple hypothesis explanations and rankings for the given descriptions or questions themselves. We analyzed the effectiveness of retrieval enhancement by comparing the hypothesis ranking of answers between the ablation group and the control group.

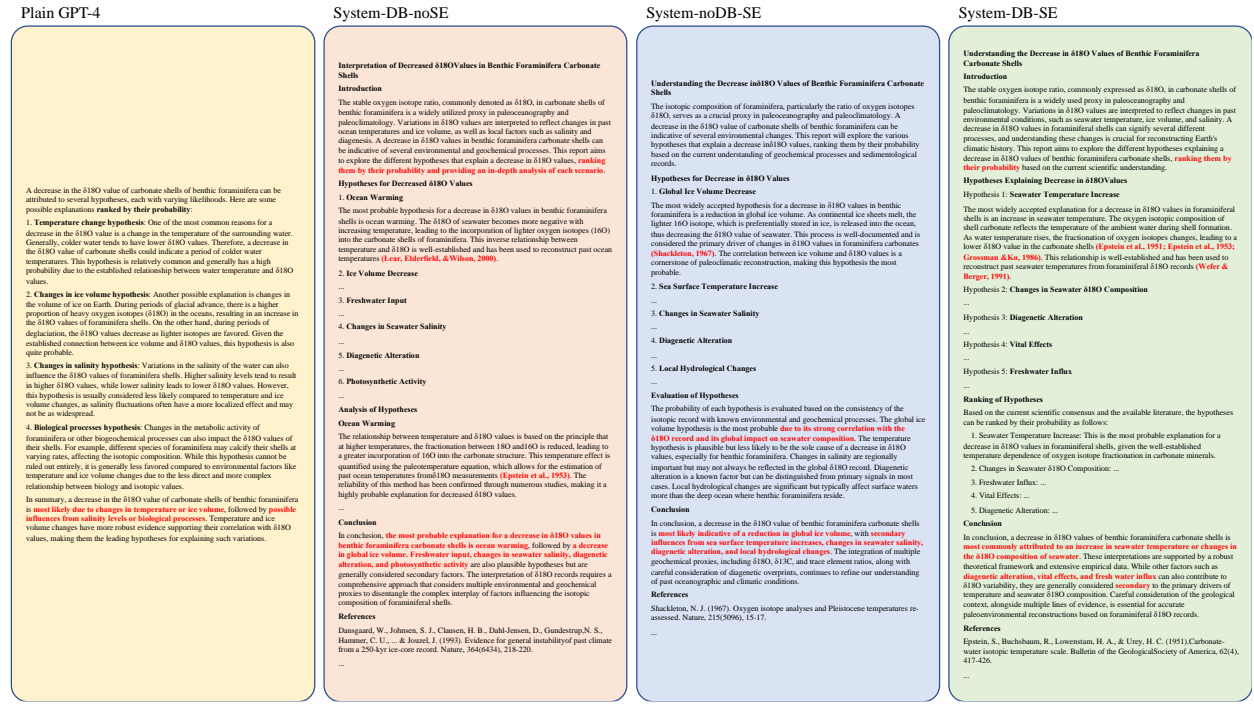

Figure 3. Comparison of ablation experiment results. They are respectively plain GPT-4, Expert system that only searches databases (System-DB-noSE), Expert system that only searches web pages (System-noDB-SE), and Expert system that searches databases and web pages (complete expert system, System-DB-SE).

We leverage GPT-4 for LLM selection. As depicted in Figure 3, employing questions and answers crafted by the expert system results in significantly enriched content compared to direct outputs from GPT-4. This improvement stems from the formulation of proper question prompts within the system, which stimulate the generation of multiple hypotheses subsequently ranked based on their probability. Consequently, all four groups of answers satisfy the specified questioning criteria.

Regarding writing style and citation practices, the effectiveness of the last three responses is attributed to the proper prompt, where these responses draw upon retrieved references. As illustrated in Figure 3, the comprehensive expert system demonstrates a greater number and density of cited references compared to the ablation group. Notably, the hypotheses proposed by all three systems demonstrate scientific validity. Additionally, the quality of question answering indicates that expert systems utilizing databases and online data retrieved from web pages outperform others.

From a logical mechanism standpoint, the system incorporates CoT. CoT enhances model capability in comprehending and generating coherent text by integrating coherence and logic tasks. Specifically, CoT operates through a series of logic-based subprocesses for generation. This approach necessitates strict logical coherence between model-generated answers and posed questions. Consequently, this training regimen directs the model's attention towards context and logical coherence, thereby improving the consistency and coherence of generated responses.

Moreover, the system utilizes RAG, which integrates retrieval and generation methodologies within its mechanism, prioritizing the extraction of information from relevant knowledge during answer generation. Initially, expert system employs a retrieval module to gather relevant documents or passages based on input questions. Subsequently, the generation module utilizes this retrieved knowledge in conjunction with the posed question to formulate answers. This mechanism facilitates the integration of external knowledge, thereby providing more precise and comprehensive information support. By incorporating retrieval knowledge, system achieves higher accuracy and content richness in generated responses.

In summary, the combination of proper prompt, CoT, and RAG enhances response logic and consistency by training models to produce coherent text while leveraging retrieval knowledge to deliver more accurate and content-rich answers. This synergy equips large models with an improved understanding of context, consideration of logical relationships, and integration of external knowledge, thereby enhancing the quality of generated responses.

**CONCLUSION**

Our study introduces an innovative approach that leverages LLMs along with retrieval augmented generation and real-time search capabilities to counteract interpretation biases. Through ablation experiments, we have substantiated notable enhancements in response quality. The integration of LLMs, chain-of-thought, contextually appropriate prompts tailored to our objectives, and retrieval-augmented generation surpasses the efficacy of employing standalone LLMs. Moreover, we emphasize that the quality of responses is intricately linked to the accessibility of pertinent data during retrieval processes. The potential of this approach is profound, poised to revolutionize our research methodologies. Our research illuminates the transformative potential of LLMs in refining paleoenvironmental studies and extends their applicability across various sub-disciplines of Earth sciences, enabling a deeper and more accurate depiction of Earth's evolution.